\documentclass[aps,twocolumn,prl,superscriptaddress]{revtex4-2}
\usepackage[tbtags,fleqn]{amsmath}
\usepackage{amssymb}
\usepackage{amsfonts}
\usepackage{mathtools}
\usepackage[colorlinks=true,citecolor=blue,urlcolor=blue]{hyperref}
\hypersetup{colorlinks,linkcolor=blue,filecolor=blue,urlcolor=blue,citecolor=blue}
\usepackage{multirow}
\usepackage{graphicx}
\usepackage{subfigure}
\usepackage{cases}
\usepackage{ulem}
\usepackage{xr}

\usepackage{appendix}

\usepackage{times}

\begin{document}
	\title{
		Stable and High-Precision 3D Positioning via Tunable Composite-Dimensional Hong-Ou-Mandel Interference}
	
	\author{Yongqiang Li}
	\affiliation{Key Laboratory of Low-Dimension Quantum Structures and Quantum Control of Ministry of Education, Synergetic Innovation Center for Quantum Effects and Applications, Xiangjiang-Laboratory and Department of Physics, Hunan Normal University, Changsha 410081, China}
	\affiliation{Institute of Interdisciplinary Studies, Hunan Normal University, Changsha 410081, China}
	
	\author{Hongfeng Liu}
	\affiliation{Department of Physics, State Key Laboratory of Quantum Functional Materials, and Guangdong Basic Research Center of Excellence for Quantum Science, Southern University of Science and Technology, Shenzhen 518055, China}
	
	\author{Dawei Lu}\thanks{Corresponding author:ludw@sustech.edu.cn}
	\affiliation{Department of Physics, State Key Laboratory of Quantum Functional Materials, and Guangdong Basic Research Center of Excellence for Quantum Science, Southern University of Science and Technology, Shenzhen 518055, China}
	
	\author{Changliang Ren}\thanks{Corresponding author: renchangliang@hunnu.edu.cn}
	\affiliation{Key Laboratory of Low-Dimension Quantum Structures and Quantum Control of Ministry of Education, Synergetic Innovation Center for Quantum Effects and Applications, Xiangjiang-Laboratory and Department of Physics, Hunan Normal University, Changsha 410081, China}
	\affiliation{Institute of Interdisciplinary Studies, Hunan Normal University, Changsha 410081, China}
	
	\begin{abstract}
		We propose a stable and high-precision three-dimensional (3D) quantum positioning scheme based on Hong-Ou-Mandel interference. While previous studies have explored HOM interference in quantum metrology, they were mostly limited to one-dimensional scenarios, whereas real-world applications require full 3D spatial resolution. Our approach not only generalizes HOM positioning to 3D—achieving ultimate sensitivity as defined by the quantum Cramér-Rao bound—but also stabilizes estimation accuracy through simple polarization tuning, ensuring that the Fisher information remains independent of the estimated parameters. Theoretical analysis and simulations demonstrate that our method achieves ultra-precise and reliable 3D positioning, even with a limited number of detected photons.

	\end{abstract}

	
	\maketitle

	\textit{Introduction---}Photons, as fundamental carriers of quantum information, possess a variety of degrees of freedom—including frequency, polarization, and momentum—that have enabled groundbreaking advancements in quantum communication \cite{Duan-Nature-2001,gisin-2007-Nat-Photonics,chen-Nature-2021}, computing \cite{bennett-2000-nature,nielsen-2010}, and metrology \cite{Giovannetti-2006-PRL,szczykulska-2016-Adv-Phys-X,Giovannetti-2011-Naturephotonmics,Barbieri-2022-PRXQuantum,Joo-2011-PRL}. By utilizing quantum superposition and interference, quantum metrology can surpass classical limits, achieving ultimate precision as dictated by the Quantum Cramér–Rao Bound (QCRB) \cite{liu-2019-J-Phys-A-Math-Theor,paris2009quantum}.
	
	Quantum positioning, an emerging frontier in quantum metrology, has witnessed significant progress in distance measurement via photon time-of-flight \cite{zhuang-2022-PRL,giovannetti-2001-Nature,Zhuang-2021-PRL} and radial speed estimation using the Doppler effect \cite{zhuang-2024-PRL,reichert-2022-quantum,Huang-2021-PRXQuantum}. However, real-world target positioning requires precise information in all three spatial dimensions. Conventional methods relying on time-delay measurements provide quantum enhancement only along the radial axis, limiting their applicability to full three-dimensional (3D) localization. Fortunately, the rich degrees of freedom of photons offer new pathways to improve lateral positioning precision, enabling the extension of quantum parameter estimation to higher-dimensional scenarios.
	
	Early quantum positioning and parameter estimation strategies typically relied on orthogonal squeezing \cite{Xiao-1987-PRL,Giovannetti-2004-Science,Caves-1981-PhysRevD,Pezze-2008-PRL,Lang-2013-PRL} or N00N-state interferometry \cite{Muller-2017-PRL,Rozema-2014-PRL,Kacprowicz-2010-Nat.Photonics,Bouwmeester-2004-Nature,kim-2016-Sci.-Reps} to enhance resolution. However, these approaches demand extremely high phase stability in noisy environments, restricting their practicality \cite{Weedbrook-2012-RevModPhys,Jordan-2022-PRA}. In contrast, Hong–Ou–Mandel (HOM) interference \cite{Hong-Ou-Mandel-1987-PRL}, a fundamental quantum interference phenomenon, is highly sensitive to group delay while being robust against phase fluctuations \cite{Hong-Ou-Mandel-1987-PRL,Bouchard-2021-Rep.Prog.Phys.,Jordan-2022-PhysRevA} and certain types of dispersion \cite{Steinberg-1992-PhysRevA,Mazzotta-2016-PhysRevA,Giovannetti-2001-PRL}. This makes HOM interference particularly advantageous for precision measurement in realistic, noisy settings. A wealth of studies has demonstrated that HOM interference excels in ultra-precise time delay measurements \cite{Ashley-2018-Sci.Adv.,Chen-2019-npj,Scott-2020-PRA,Mandel-1999-RevModPhys,Triggiani-2023-Phys.-Rev.-Appl.}, and recent research further confirms its ability to achieve quantum-limited sensitivity in transverse displacement estimation \cite{Triggiani-2024-PRL,Triggiani-2025-PRA,Triggiani-frequency-shift-2024-arxiv,Triggiani-incoherent-2024-arxiv}. Interestingly, although the potential of the HOM effect for precision metrology has been recognized for decades \cite{Weedbrook-2012-RevModPhys}, a rigorous analysis of the ultimate precision limits of HOM-based measurements has only been pursued in recent years \cite{Jordan-2022-PhysRevA,Triggiani-2024-PRL}. Moreover, the existing literature has largely overlooked how the estimation accuracy of HOM-based schemes is affected by the target parameters. Ensuring that measurement precision remains independent of these parameters is essential for the stability and practical implementation of quantum metrology.

	Motivated by these challenges, we propose a stable and high-precision 3D quantum positioning strategy that exploits the unique influence of various photonic degrees of freedom on HOM interference. Unlike previous schemes limited to one-dimensional radial or transverse measurements, our method incorporates both transverse momentum and frequency to achieve comprehensive 3D spatial localization. Most significantly, by fine-tuning polarization detection, we demonstrate that our method not only ensures positioning accuracy independent of the target parameters but also reaches the ultimate sensitivity imposed by the QCRB. Theoretical analysis further confirms that our strategy can achieve high-precision 3D positioning with minimal photon detection, making it highly practical for quantum metrology applications.

	\begin{figure}[htbp]
		\centering
		\includegraphics[width=0.42\textwidth,height=0.42\textwidth]{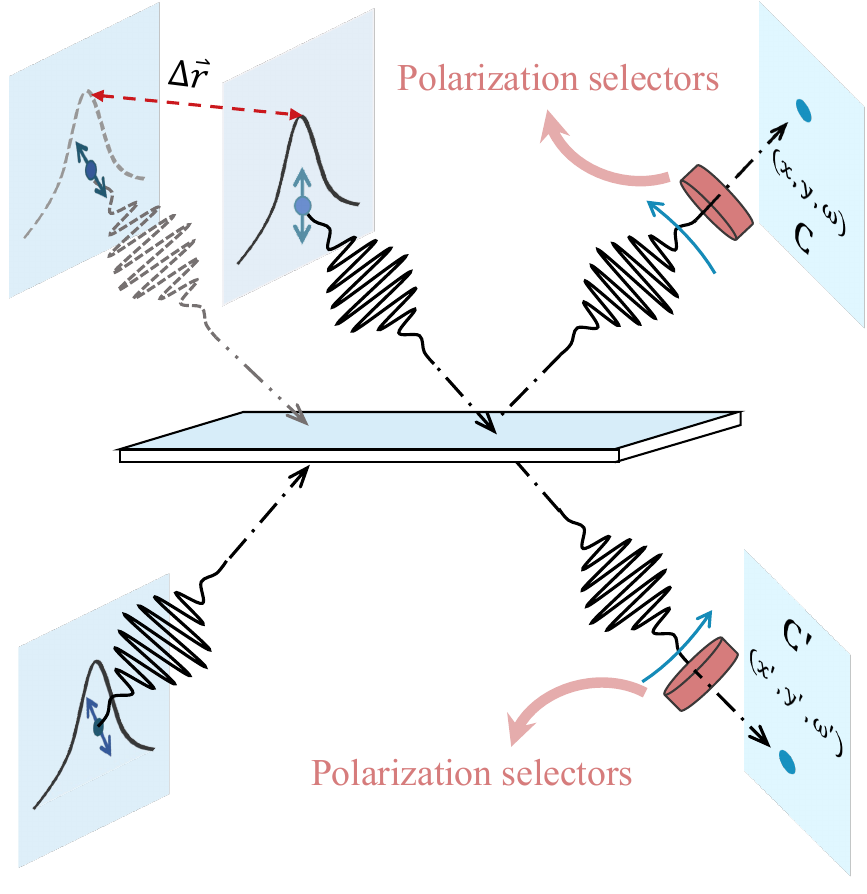}\vspace{-0.4cm}
		\caption{
			Conceptual schematic of 3D positioning via polarization-tuned HOM interference. Photons from sources at  $(\bar{\rho}_{1},d_1)$ and $(\bar{\rho}_{2},d_2)$ impinge on a balanced beam splitter, with polarization selectors at the output ports enabling projective measurements. The photons are then detected by cameras (\(C\) and \(C^{\prime}\)) in the far-field regime.}\label{figure 1}\vspace{-0.5cm}
	\end{figure}
	
	\textit{ Scheme of Estimating 3D spatial position of photons.---} \label{II}
	Consider the depiction in Fig.~\ref{figure 1} of two independent photon sources which are located the both side of the beam splitter, the task involves using one source as a reference to accurately measure and estimate the transverse and radial position offsets of the other photon source. We assume that the position of the two  single-photon source are $\vec{r}_{j}=(\vec{\rho}_{j},d_{j})$ respectively, with $j=\{1,2\}$, where $\vec{\rho}_{j}=(x_{j},y_{j})$ is the two-dimensional transverse position and $d_{j}$ is the radial position. Obviously, the position offsets between of them is $\Delta \vec{r}=\vec{r}_{1}-\vec{r}_{2}=(\Delta \vec{\rho},\Delta d)$, where $\vec{\rho}=\vec{\rho}_{1}-\vec{\rho}_{2}$ and $\Delta d=d_1-d_2$. The input photon source can be described as
	\begin{eqnarray}
		|\psi\rangle_{j}=\int  d\vec{\rho}_{j}dt_{j}\phi_{j}(\vec{\rho}_{j},t_{j})\hat{a}_{j}^{\dagger}(\vec{\rho}_{j},t_{j})|0\rangle, \nonumber \label{eq.1}
	\end{eqnarray}
	where the wavepacket $\phi_{j}(\vec{\rho}_{j},t_{j})$ centred around the transverse position $\bar{\rho}_{j}$ and times of flight $\bar{t}_{j}$. $\hat{a}_{1}^{\dagger}(\vec{\rho}_{1},t_{1})$ and $\hat{a}_{2}^{\dagger}(\vec{\rho}_{2},t_{2})$ are the bosonic creation operators associated with the first and second input modes of the beam splitter. In addition to the transverse position and time degrees of freedom, the modes of the two sources also depend on distinct polarization properties. Without loss of generality, assuming that $\hat{a}_{1}(\vec{\rho}_{1},t_{1})$ represents the signal to be measured and $\hat{a}_{2}(\vec{\rho}_{2},t_{2})$ is the reference signal, we define 
	$\hat{a}_{1}(\vec{\rho}_{1},t_{1})=\sqrt{\nu}\hat{a}_{1,H}(\vec{\rho}_{1},t_{1})+\sqrt{1-\nu}\hat{a}_{1,V}(\vec{\rho}_{1},t_{1})$ 
	and $\hat{a}_{2}(\vec{\rho}_{2},t_{2})=\hat{a}_{2,H}(\vec{\rho}_{2},t_{2})$ respectively, where $\{\hat{a}_{j,H},\hat{a}_{j,V}\}$ indicate the orthogonal polarization basis vectors of the modes. So the commutation relation satisfies $[\hat{a}_{i}(\vec{\rho}_{1},t_{1}),\hat{a}_{j}^{\dagger}(\vec{\rho}_{2},t_{2})]=\sqrt{\nu}\delta_{ij}\delta(\vec{\rho}_{1}-\vec{\rho}_{2})\delta(t_{1}-t_{2})$. $\nu$ represents the degree of indistinguishability between the two photons, ranging from 0 to 1. Only when $\nu=1$,the two photons become completely indistinguishable.

	\begin{figure*}[t!]
		\centering
		\includegraphics[width=0.9\textwidth]{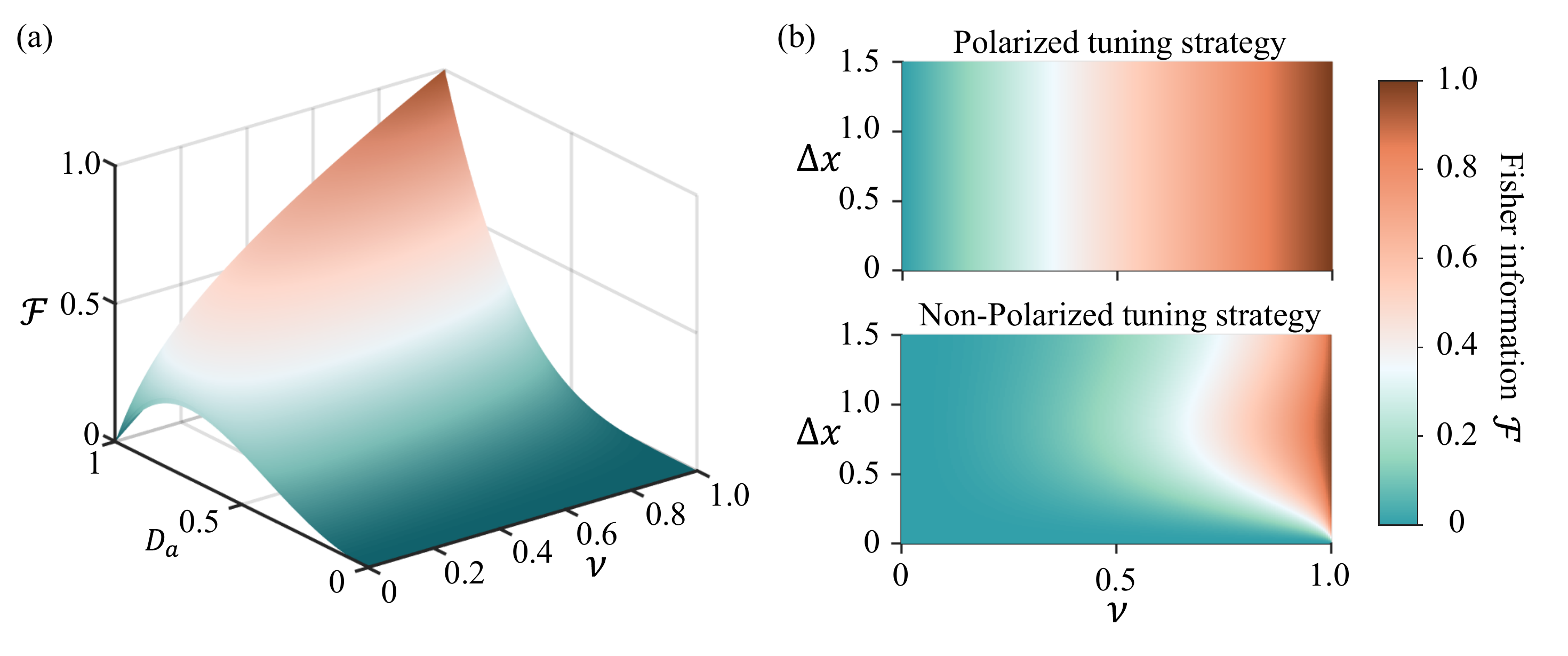}\vspace{-0.5cm}
		\caption{
			Plots of Fisher information for the polarization-tuning and non-polarization-tuning strategies. (a) Fisher information as a function of \(D_a\) and \(\nu\) for the polarization-tuning strategy. (b) Fisher information as a function of \(\nu\) and \(\Delta x\). The upper panel corresponds to the polarization-tuning strategy with \(D_a=1\), while the lower panel represents the non-polarization tuning strategy.}
		\label{figure 2}\vspace{-0.5cm}
	\end{figure*}

	As schematically illustrated in Fig.~\ref{figure 1}, after the two photons strike the two surfaces of the balanced beam splitter (BS), interference occurs under certain conditions (e.g., spectral indistinguishability and spatial mode overlap). Polarization selectors, such as polarizers, are placed at both output ports of the beam splitter to perform projective measurements on the polarization states of the outgoing photons. Finally, the signals are detected by two detectors  (\(C\) and \(C^{\prime}\)) placed at the respective output ports, which record the frequency and transverse position of the photons. Assume that detector \(C\)  ( \(C^{\prime}\)) records a photon signal with a transverse position at $\vec{\rho}$ ($\vec{\rho}^{\prime}$) and frequency $\omega$($\omega^{\prime}$). The second-order correlation function of detecting two photons can be expressed as 
	\begin{equation}
		\begin{aligned}
			&G^{\alpha, \beta}_{CC^{\prime}}(\vec{\rho},\vec{\rho}^{\prime},\omega,\omega^{\prime})= \\ & \langle\psi| \hat{E}_{C}(\vec{\rho},\omega)\hat{E}_{C^{'}}(\vec{\rho}^{\prime},\omega^{\prime})\hat{E}_{C^{'}}^{+}(\vec{\rho}^{\prime},\omega^{\prime})\hat{E}_{C}^{+}(\vec{\rho},\omega)|\psi\rangle, \label{2} 
		\end{aligned}
	\end{equation}	 
	where $\alpha$ ($\beta$) depends on the projection of the polarization before detection.  The field operators can be defined by $\hat{E}_{C}^{+}(\vec{\rho},\omega)=\sum_{j=1,2}\int dt_j d\vec{\rho}_{j} g(\vec{\rho}_{j},j;\vec{\rho},C,\alpha)e^{i\omega(t_j-\bar{t}_{j})}\hat{a}_{j}(\vec{\rho}_{j},t_j)$, where $g(\vec{\rho}_{j},j;\vec{\rho},C,\alpha)=g(\vec{\rho}_{j},j;\vec{\rho},C)P(\alpha)$, which can be divided into two terms, the Fraunhofer transfer function and polarization post-selection. So we can define an annihilation operator $\hat{\alpha}_{j}(\vec{\rho}_{j},t_j)=P(\alpha)\hat{a}_{j}(\vec{\rho}_{j},t_j)=C_a\hat{a}_{j,H}(\vec{\rho}_{j},t_j)+C_b\hat{a}_{j,V}(\vec{\rho}_{j},t_j)$. By tuning the polarizer, the distribution of the orthogonal modes of the annihilation operator is continuously altered. The operator $\hat{E}_{C^{\prime}}^{+}(\vec{\rho}^{\prime},\omega^{\prime})$ can be similarly obtained, and it corresponds to the annihilation operator $\hat{\beta}_{j}(\vec{\rho}_{j},t_j)=P(\beta)\hat{a}_{j}(\vec{\rho}_{j},t_j)=D_a\hat{a}_{j,H}(\vec{\rho}_{j},t_j)+D_b\hat{a}_{j,V}(\vec{\rho}_{j},t_j)$. 
	
	With a small radial offset, the second-order correlation function is analytically determined as a parametric function of transverse momentum variation $\Delta \vec{k}$ and frequency difference $\Delta \omega$, 
	\begin{equation}
		\begin{aligned}
			\mathrm{G}(\Delta \vec{k}, \Delta \omega)=\Gamma+\Upsilon  \Theta \cos(\Delta \vec{k}\cdot \Delta \bar{\rho}+\Delta \omega \Delta \bar{t}),\label{eq.2}
		\end{aligned}
	\end{equation}
	where $\Delta \vec{k}=\vec{k}-\vec{k}^{\prime}=(\Delta k_{x},\Delta k_{y})$, with $\Delta k_{x}=\frac{k_{0}}{d}x-\frac{k_{0}}{d}x^{\prime}$ and $\Delta k_{y}=\frac{k_{0}}{d}y-\frac{k_{0}}{d}y^{\prime}$. The $\Delta \omega=\omega-\omega^{\prime}$ is the frequency different between the two photons. $\Upsilon$ is used to clarify whether two photons are detected by different detectors or if they are both recorded by the same detector, where $\Upsilon=\pm 
	1$. $\Gamma$ and $\Theta$ are determined by three key factors, spectral distribution characteristics $\mathrm{\phi}$,  photon distinguishability parameter $\mathrm{\nu}$, and polarization post-selection of the detection events. The detailed derivation formula is rigorously presented in the Appendix (I.A).

	Eq.~\eqref{eq.2} clearly demonstrates that the statistical distribution of measurement results enables the estimation of both transverse position offsets ($\Delta \bar{\rho}=(\Delta x, \Delta y)$) and time delay ($\Delta \bar{t}$) between the two photon sources. The estimation accuracy for every parameter is fundamentally constrained by the Fisher information which gives the theoretical lower bound. The relationship is mathematically expressed through the Cramér-Rao inequality \cite{paris2009quantum},
	\begin{eqnarray}
		\text{Var}(\Delta \theta) \geq \frac{1}{N \mathcal{F}[\Delta \theta]},
	\end{eqnarray}
	where $\Delta \theta$ denotes the parameter vector to be estimated and $N$ represents the number of measurements. The Fisher information metric for these parameters is formally defined as $\mathcal{F}[\Delta\theta] = \mathbb{E}\left[ \left( \frac{\mathrm{d}}{\mathrm{d}\Delta\theta} \log (\mathrm{G}(\Delta \vec{k}, \Delta \omega)) \right)^2 \right]$,
	where $\mathbb{E}[\cdot]$ corresponds to the statistical expectation operator. This formulation quantifies the sensitivity of the probability distribution $\mathrm{G}(\Delta \vec{k}, \Delta \omega)$ to parameter variations, thereby determining the ultimate precision limit in quantum parameter estimation.       
	
	By substituting Eq.~\eqref{eq.2} into the Fisher information formula, we arrive at 
	\begin{eqnarray} 
		\mathcal{F}[\Delta \theta] = \int \mathrm{d}^{2}\Delta \vec{k} \, \mathrm{d}\Delta \omega \, f(\Delta \vec{k}, \Delta \omega, \Delta \theta),
	\end{eqnarray} 
	with $f(\Delta \vec{k},\Delta \omega, \Delta \theta)=\mathrm{G}(\Delta \vec{k},\Delta \omega)(\frac{\mathrm{d}}{\mathrm{d} \theta}\log(\mathrm{G}(\Delta \vec{k},\Delta \omega)))^2$, where $\mathrm{G}(\Delta \vec{k},\Delta \omega)$ is the joint probability distribution for detecting two photons. The exact analytical form of $\mathcal{F}[\Delta \theta]$ is provided in the equations (\ref{eq.B.16}-\ref{eq.B.18}) within the Appendix. The results clearly indicate that $\mathcal{F}[\Delta \theta]$ depends on the estimated parameter $\Delta \theta$. Nonetheless, this problem can be solved by appropriate polarization tuning, the Fisher information will be independent of any estimated parameters $\Delta \theta$ as long as the following condition is met, $2C_{a}D_{a}C_{b}D_{b}=C_{a}^{2}D_{b}^{2}+C_{b}^{2}D_{a}^{2}$, which implies $\Gamma = \Theta$ in Eq.~\eqref{eq.2}. Achieving this condition requires minimal effort in experiment. For example, it can be satisfied by setting $C_{a}=D_{a}$ and $C_{b}=D_{b}$, which means $\hat{\alpha}_j(\vec{\rho}_{j},t_j)=\hat{\beta}_j(\vec{\rho}_{j},t_j)$. In this case, the interference fringe visibility can always be maintained at its maximum , $\mathcal{I}=1$.

	Without loss of generality, supposed that the spectral distribution is Gaussian, $\phi(\Delta \vec{k},\Delta \omega)\propto e^{-\frac{\Delta k_{x}^2}{4\sigma_{k_{x}}^2}-\frac{\Delta k_{y}^2}{4\sigma_{k_{y}}^2}-\frac{\Delta \omega^2}{4\sigma_{\omega}^2}}$, where $\phi(\Delta \vec{k},\Delta \omega)$ is the Fourier transformation of $\phi(\Delta \vec{\rho},\Delta t)$. The Fisher information matrix can be given as
	\begin{eqnarray}
		\mathcal{F}=\gamma\begin{pmatrix}
			\sigma_{k_{x}}^2 & 0 &0 \\
			0&  \sigma_{k_{y}}^2 &0 \\
			0& 0 & \sigma_{\omega}^2
		\end{pmatrix}, \label{QFIM}
	\end{eqnarray}
	with $\gamma=D_{a}^2(D_{a}\sqrt{\nu}+D_{b}\sqrt{1-\nu})^2$, $D_{b}=\sqrt{1-D_{a}^2}$. The precision of the parameters is inversely proportional to the corresponding elements of this matrix. Since the Fisher information matrix is a diagonal matrix, the transverse position and time can be estimated simultaneously. In principle, the accuracy can be improved by increasing the bandwidth of transverse-momentum and frequency. Additionally, for different parameters $\nu$, by tuning the polarization post-selection (i.e., changing $D_{a}$), the coefficient $\gamma$ can be increased, thus further optimizing the Fisher information. For the fixed $\nu$, the optimal $D_{a}$ is $D_{a} = \frac{\sqrt{1 + \sqrt{\nu}}}{\sqrt{2}}$, which leads to the maximal $\gamma = \frac{(1 + \sqrt{\nu})^2}{4}$. Fig.\ref{figure 2}(a) shows the dependence of the Fisher information on photon indistinguishability $\nu$ and polarization parameter $D_{a}$. Taking $\Delta x$ as an example, it reveals that as $\nu$ and $D_{a}$ increase, the strategy enhances the accuracy of the estimation of transverse position and flight time. When $D_{a}$ and $\nu$ approach $1$, the maximum accuracy by this strategy achieved, which arrives the maximum Fisher information allowed by the photon spectral distribution.
	
\begin{figure}[t!]
		\centering
		\includegraphics[width=0.45\textwidth,height=0.36\textwidth]{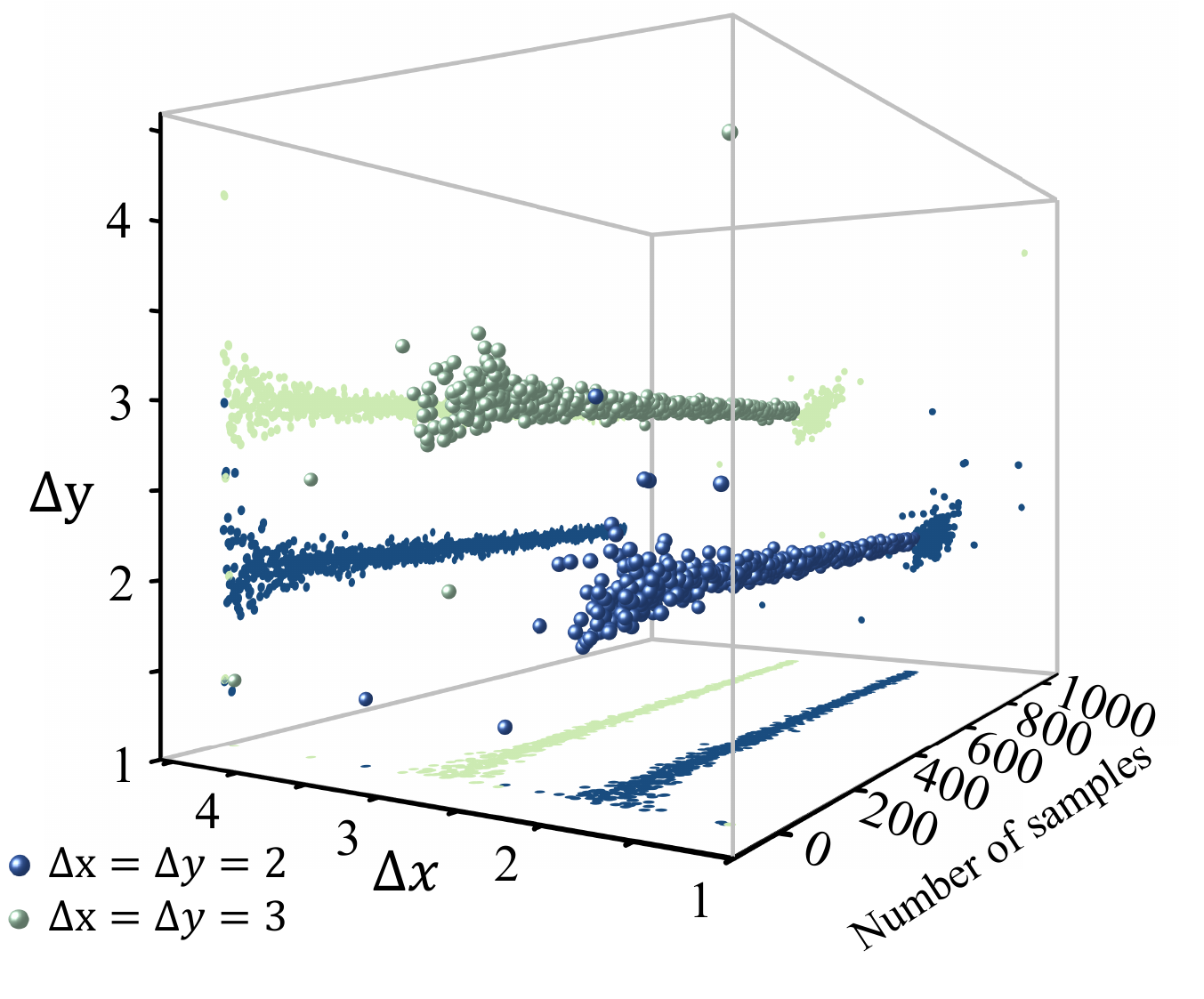}\vspace{-0.4cm}
		\caption{Numerical simulations for different values \(\Delta \bar{\rho}\) and \(\phi(\Delta \vec{k}, \Delta \omega)\) with \(\sigma_{k_{x}}=\sigma_{k_{y}}=1\) and \(\gamma=\frac{1}{2}\). The blue and green markers indicate the cases of the $\Delta x=\Delta y=2$ and $\Delta x=\Delta y=3$, respectively. The accuracy of the estimators for $\Delta x$ and $\Delta y$ improves with increasing number of samples.
		}\label{figure 3}\vspace{-0.5cm}
	\end{figure}

	We perform a numerical analysis of the scheme's resource consumption. With the parameters set to $D_a = 1/\sqrt{2}$ and \(\nu = 1/2\), the derived parameter \(\gamma\) consequently equals \(1/2\). Under the assumption of Gaussian distributions \(\phi(\vec{\rho}_j,t_j)\) characterized by spatial-temporal widths \(\sigma_x = 50\ \text{nm}\), \(\sigma_y = 100\ \text{nm}\), and \(\sigma_t = 0.3\ \text{fs}\), our calculations demonstrate that utilizing 1,000 photon pairs achieves remarkable estimation accuracies: \(2.2\ \text{nm}\) for \(\Delta x\), \(4.5\ \text{nm}\) for \(\Delta y\), and \(13.4\ \text{as}\) for \(\Delta \bar{t}\). Through maximum likelihood estimation simulations, We quantitatively illustrate the variation in the estimates of \(\Delta x\) and \(\Delta y\), as a function of the number of samples. As shown in Fig.~\ref{figure 3}, these parameters exhibit progressive convergence with increasing sample size. Notably, the elliptical contour observed in the \(\mathrm{\Delta x-}\mathrm{\Delta y}\) plane projection originates from the structural characteristics of  second-order correlation function.
	Notably, with current photon sources achieving generation rates of 1,000 photon pairs per millisecond, this method is not only theoretically sound but also readily implementable with existing technology.
	
	
	\textit{Comparison with the Non-polarized tuning strategy.---} 
	Recently, Triggiani \textit{et al.} proposed a scheme for estimating transverse displacement between photons via two-photon interference sampling measurements \cite{Triggiani-2024-PRL}. This approach achieves quantum-limited sensitivity and overcomes the diffraction limitation inherent in classical microscopy imaging. Until now, their work currently focuses on one-dimensional transverse displacement estimation \cite{Triggiani-2025-PRA,Triggiani-incoherent-2024-arxiv}. While they propose that the method could, in principle, be adapted for full 3D localization, realizing this extension requires overcoming non-trivial technical hurdles. We develop an adapted version of this strategy for 3D positioning estimation and conduct a detailed comparison with our scheme. The experimental configuration closely mirrors the schematic shown in Fig.~\ref{figure 1}, though differs through the polarization detection. Comprehensive theoretical derivations and technical analysis of this quantum-enhanced displacement estimation framework are elaborated in Appendix~(I.B).  
	
	Assumed that the same photon source defined in equation~\eqref{eq.1} is used, the probability distribution of photons registered by the detectors derived as a parametric function of transverse momentum variation $\Delta \vec{k}$ and frequency difference $\Delta \omega$  (\ref{eq.C.12}),
	\begin{eqnarray}
		\mathrm{P}=\frac{1}{2}| \phi(\Delta \vec{k}, \Delta \omega) |^2  (1 + \nu \Upsilon \cos(\Delta \vec{k} \cdot\Delta \bar{\rho} + \Delta \omega \Delta \bar{t}) ).
	\end{eqnarray}
	Similarly, the Fisher information for each the parameter estimation $\Delta \theta$ can be derived,
	\begin{eqnarray}
		\mathcal{F}^{\prime}[\Delta \theta]= \int \mathrm{d}^2\Delta \vec{k}\mathrm{d}\Delta \omega f^{\prime}(\Delta \vec{k},\Delta \omega,\Delta \theta) \label{eq.8}
	\end{eqnarray}
	where $f^{\prime}(\Delta \vec{k},\Delta \omega,\Delta \theta)$ is a function that depends on $\Delta \theta$. And the exact analytical form of $\mathcal{F}^{\prime}[\Delta \theta]$ is provided in the Eqs. (\ref{eq.B.43}-\ref{eq.B.45}) within the Appendix. Thus, within this strategy, the Fisher information matrix corresponding to parameter estimates intrinsically depends on the parameter values, leading to variations as the parameters change. The measurement accuracy of their strategy remains independent of these parameters only when complete photon indistinguishability is ensured. In contrast, our scheme consistently guarantees that the estimation accuracy remains independent of the parameters.Fig.\ref{figure 2}(b) illustrates the relationship between Fisher information and the parameter for the two different schemes (the upside is our scheme). Similarly, we analyze the interference visibility in both schemes. Obviously, the polarization-tuning scheme consistently maintains a visibility of \(\mathcal{I}=1\), whereas the visibility in the other scheme is contingent on the degree of similarity between the two photon modes \(\mathcal{I}=\nu\).
	
	\begin{figure}[t]
		\centering
		\includegraphics[width=1\linewidth]{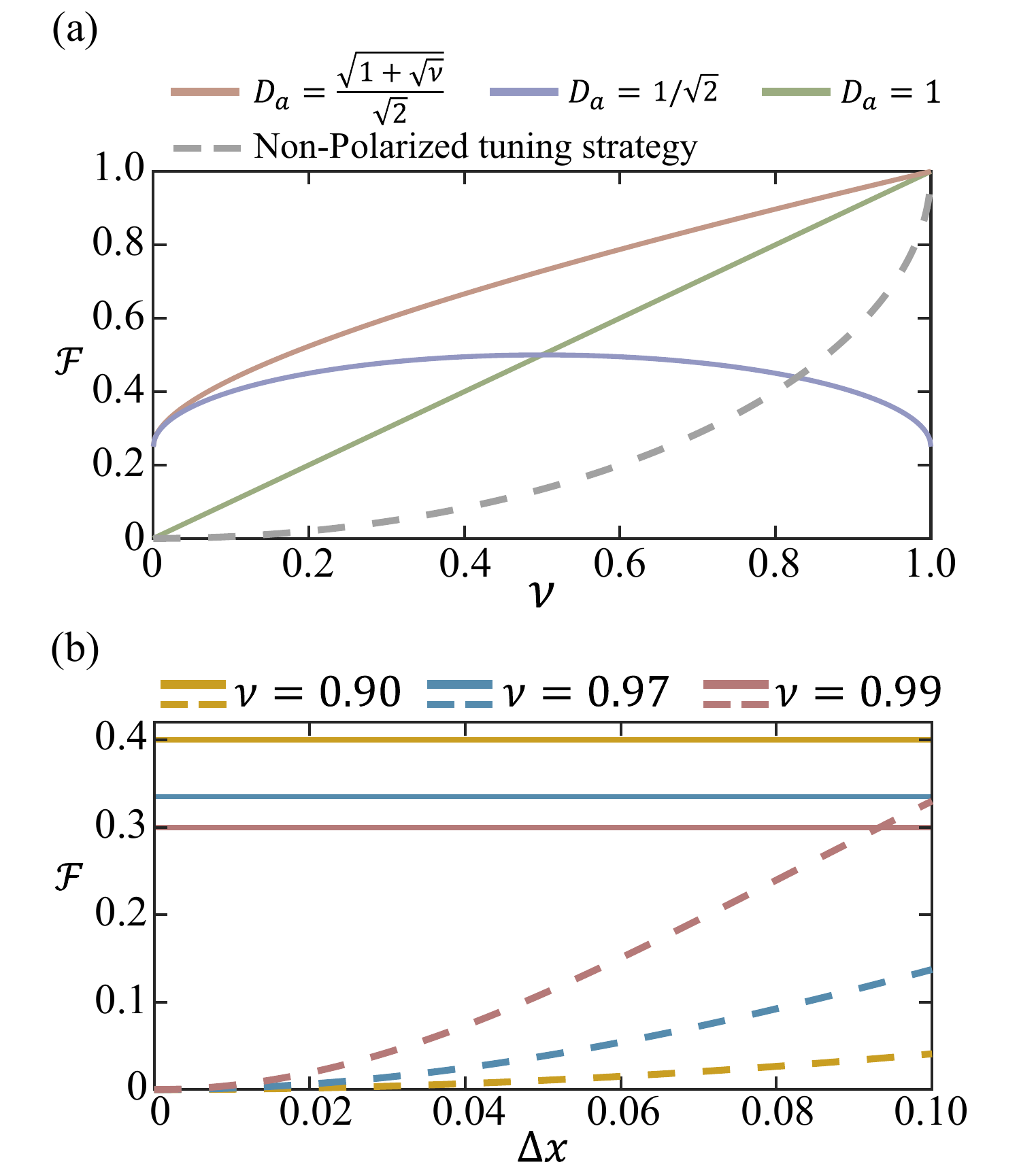}\vspace{-0.3cm}
		\caption{Comparison of Fisher information for polarization-tuning and non-polarization-tuning strategies. (a) Fisher information as a function of \(\nu\), where solid lines correspond to the polarization-tuning strategy with distinct values of \(D_a\) indicated by different colors, and the dashed line represents the non-polarization-tuning strategy. (b) Fisher information versus the estimated parameter, where the solid line represents the polarization-tuning strategy with \(D_a=\frac{1}{\sqrt{2}}\), and different colors indicating various \(\nu\). Likewise, the dashed line represents for the non-polarization tuning strategy.
		}\label{figure 4}\vspace{-0.5cm}
	\end{figure}

	The detailed analysis of the accuracy of parameter estimation based on this non-polarized tuning scheme has been derived in Appendix A.2. Although the parameter estimation accuracy in this scheme depends on parameter itself, while under specific conditions (\(\sigma_{k_{x}} \Delta \bar{x} \gg 1\), \(\sigma_{k_{y}} \Delta \bar{y} \gg 1\) and  \(\sigma_{\omega} \Delta \bar{t} \gg 1\)), it can be approximated as independent of the estimated parameters. The Fisher information matrix~\eqref{eq.B.48} remains similar to  Eq.~\eqref{QFIM}, except that the scaling coefficient $\gamma$ is replaced by $\kappa$ ($\kappa=1-\sqrt{1-\nu^2}$).
	Without loss of generality, we set all bandwidths to natural constants. So the Fisher information of the estimated parameter is analytically derived as a function of photon indistinguishability coefficient  \(\nu\), following Eq.~\eqref{eq.B.48}. As illustrated in Fig.~\ref{figure 4}(a), the red dotted line represents the corresponding Fisher information based on the non-polarized tuning scheme, exhibiting a monotonic increase with \(\nu\) and reaching its theoretical maximum at \(\nu=1\), where the measurement precision of HOM interferometry saturates the fundamental limit. For comparison, the Fisher information for parameter estimation based on our scheme was also analytically obtained and plotted under the same parameter conditions. The red solid line Fig.~\ref{figure 4}(a) represents the maximum Fisher information achievable based on our scheme, which corresponds to \(D_a=\frac{\sqrt{1+\sqrt{\nu}}}{\sqrt{2}}\).
	Clearly, the Fisher information achieved through our approach consistently exceeds that of the non-polarized tuning scheme. Even when tuning is not optimal, the fisher information can exceed that of the non-polarized tuning scheme in most regions, such as the green  (\(D_a=\frac{1}{\sqrt{2}}\)) and blue (\(D_a=1\)) solid line. In addition, even in the extreme case of completely orthogonal photon polarizations (\(\nu=0\)), the Fisher information extraction remains possible through optimized polarization tuning. This highlights the unique advantage of polarization tuning in surpassing classical limits in quantum parameter estimation. With the same conditions and detection resources, polarization tuning naturally leads to higher estimation accuracy. Such as, 
	when $\nu=\frac{1}{2}$, assuming Gaussian distributions $\phi(\vec{\rho}_{j},t_{j})$ with a width $\sigma_{x}=50\ \text{nm}$, $\sigma_{y}=100\ \text{nm}$, and $\sigma_{t}=0.3\ \text{fs}$,   
	the accuracy of estimating $\Delta x$, $\Delta y$, and $\Delta \bar{t}$ based on the non-polarized tuning scheme reaches $4.3\ \text{nm}$, $8.6\ \text{nm}$, and $25.9\ \text{as}$ when employ 1000 pairs of photons. With polarization tuning, the estimation accuracy under identical conditions can reach \(2.2\ \text{nm}\), \(4.5\ \text{nm}\), and \(13.4\ \text{as}\) respectively.
	
	For using non-resolved detectors, we analyze and derive the Fisher information under two different schemes ((\ref{B.23}-\ref{B.25}) and (\ref{B.50}-\ref{B.52})). When two photons with highly overlapping spatial wave packets (i.e. satisfying $\sigma_{k_x}\Delta x \ll1, \sigma_{k_y}\Delta y\ll1$, and $\sigma_{\omega}\Delta \bar{t}\ll1$), an approximate analytical result is obtained for both ((\ref{eq.B.27}-\ref{eq.B.29}) and (\ref{eq.C.24}-\ref{eq.C.26})). In single-parameter estimation, the polarization tuning scheme ensures parameter-independent Fisher information \eqref{eq.B.30}, while the non-tuning scheme depends on the parameter \eqref{B.56}. In multi-parameter estimation with a bucket detector, both methods exhibit parameter dependence ((\ref{eq.B.27}-\ref{eq.B.29}) and (\ref{eq.C.24}-\ref{eq.C.26})).

	\textit{Conclusion---} In this letter, we introduce a quantum positioning scheme that harnesses multiple photonic degrees of freedom to achieve robust and efficient three-dimensional localization through their distinct influence on Hong-Ou-Mandel (HOM) interference. The wave packet spacing can be precisely determined by sampling the two-dimensional transverse momentum and frequency of photons, reaching the ultimate quantum sensitivity. Notably, polarization tuning effectively modifies the dependence of the Fisher information on the estimated parameters. Unlike the non-polarized-tuning strategy, appropriate polarization tuning preserves localization precision regardless of the estimated parameters. The corresponding measurement requirements can be readily implemented with current experimental techniques. Numerical analysis reveals that high-precision 3D positioning can be achieved even with a limited number of detected photons. The comparison with non-polarization-tuned scheme demonstrates the clear benefits of our approach. These findings provide valuable guidance for the future development of practical quantum positioning techniques.

The proposed scheme is not only applicable for three-dimensional positioning, but also allows simultaneous estimation of the target’s angular velocity and transverse position by sampling the time and transverse momentum distributions of photons (see Appendix (II)). Additionally, polarization is just one of many degrees of freedom available for photons, and other freedoms can also be used in place of polarization for positioning. For instance, by utilizing a Laguerre-Gaussian source, the target's three-dimensional spatial position can be inferred by distinguishing different Laguerre-Gaussian modes. In principle, this strategy admits a straightforward generalization to systems involving multi-photon Hong-Ou-Mandel interference , providing new possibilities for enhancing positioning accuracy.

	\textit{Acknowledgment---}C.R. was supported by the National Natural Science Foundation of China (Grants No. 12075245, 12421005 and No. 12247105), Hunan provincial major sci-tech program (No. 2023ZJ1010), the Natural Science Foundation of Hunan Province (2021JJ10033), the Foundation Xiangjiang Laboratory (XJ2302001) and Xiaoxiang Scholars Program of Hunan Normal University.
\bibliography{ref.bib}

	
\end{document}